\documentclass[a4paper,10pt]{article}
\usepackage[utf8]{inputenc}
\usepackage[OT4]{fontenc}
\usepackage{graphicx}
\usepackage{listings}

\begin{document}
\title{Sitnikov problem as a source of jets}
\author{Marcin Misiak}

\date{gmuse2@gmail.com}

\maketitle
\abstract {Sitnikov problem, consisting two close binaries and a third small body is considered, leading to a rapid ejection of the small body from the binaries. This mechanism is proposed as an explanation of jets in many astrophysical systems. Choosing appropriate initial condition relativistic final velocities can be achieved.}

\section{Introduction}
Sitnikov problem \cite{sit} is a special 3-body problem in which two massive bodies follow keplerian orbits and a third, with infinitesmal mass, travel along straight line perpendicular to the primaries' plane, through their barycentre. This problem acquired considerable attention of physicists mostly in the case of oscillatory motion \cite{mos}. Here, however, we focus on unbound trajectories possible in this system and propose it as a source of jets observable in many stellar and galactic systems.\\

One of the most striking observable effect in the Universe is a production of jets in many astrophysical systems which can stretch on a large distances (even million of light years) and can posses relativistic speeds. The usual model for production of these jets is a single rotating black hole surronded by accretion disk with strong magnetic field \cite{bla}. However, also differnet models are considered \cite{gar}. Here, we propose a following mechanism: two large masses (possibly black holes) forming closed binary systems, with high eccentricity elliptical orbits, surronded by accretion disk but with no magnetic field needed. Black holes can be of stellar masses as well as supermassive in the center of the galaxies.\\

Let's consider a small particle at a distance $z_0$ from the binaries, perpendicular to their plane. Assume that the binaries have the same mass and that they orbits are ellipses with high-eccentricity. Neglect any relativistic effects and treat the problem purely Newtonian. Because of the gravitational attraction from the primaries, the particle will follow straight line along the $z$ direction. We have got here the special case of the famous restricted three-body problem which is non-integrable system. If at the time of particle crossing the center of mass of the binaries they are in their periastron position (denoted B on the Fig. 1) nothing interesting happens - the particle will just follow usually oscillatory motion. To exclude this possibility let's assume that the binaries are then not in their periastron position. In this case the situation is more interesting and choosing appropiately the position of the primaries in the moment of maximum encounter, energy can be transferred from them to the particle. We are interested in the question: what is the maximum ejection velocity of the particle from the closed binary-system? This problem is related to the gravitational assist, commonly used in Solar System exploration \cite{ale}.\\

We expect the maximum effect in the case where the binaries are coming to their periastron position just before the particle attain their center of mass. In this case, when all three bodies are minimally separated from each other and the particle is still before the center of mass point, it will get a big push forward ejecting it from the system, which will not be cancelled by later slowing down of the particle.

\section{Numerical computation}

To answer the question we performed numerical integration choosing the initial position of the binaries in their periastron position (denoted B on the diagram), the initial velocity of the particle as zero and we were changing the initial distance $z_0$ from the particle to the center of the mass of the system as to find the best initial position leading to a maximal ejection velocity of the particle along a straight line away from the system. The two masses forming binary system follow keplerian orbits and to determine their position in any time we used the Kepler equation. Our system of equation is:

\begin{eqnarray}
\frac{dz}{dt} & = & v\\ \nonumber
\frac{dv}{dt} & = & \frac{-2GMz}{(z^2+r^2)^{\frac{3}{2}}}\\ \nonumber
r & = & a\cdot(1-e\cdot cos(u))\\ \nonumber
\end{eqnarray}

where $u$ is calculated at any time $t$ from the Kepler equation \cite{wie,roy}:
\begin{equation}
\frac{2\pi t}{T} = u-e\cdot sin(u) \nonumber
\end{equation}.

The trajectory of the particle is integrated using the standard 4th order Runge-Kutta method. The code we used is presented in the appendix.\\

We assumed at first the following data of the closed binary system: semimajor axis $a=6\cdot 10^7m$, semiminor axis $b=2\cdot 10^7m$, G (gravitational constant) and M (mass of the star) product $G\cdot M=10^{22}$. We've found the maximum velocity which the particle received as $1.31\cdot 10^8\frac{m}{s}$ when crossing the plane of the binary system. Later the gravitational attraction was slowing down the particle but assymptotically (far away from the system) it still kept the velocity equal to $0.73\cdot 10^7\frac{m}{s}$ that is above $20\%$ the speed of light. This happened when the initial distance $z_0$ was equal to $-4.931\cdot a$.\\

As a second example we considered the PSR1913+16-like system and choosed the following data: semimajor axis $a=1.95\cdot 10^9m$, eccentricity $e=0.617131$, orbital period $=27900s$, GM product $G\cdot M=2\cdot 10^{20}$. In this case because of the smaller masses and larger parameters of the ellipses the effect was much smaller but still the particle attained $538\frac{km}{s}$ final velocity for the initial distance $z_0=-3.25\cdot a$.

\begin{figure}
\begin{center}
\includegraphics[width=6cm,height=7cm]{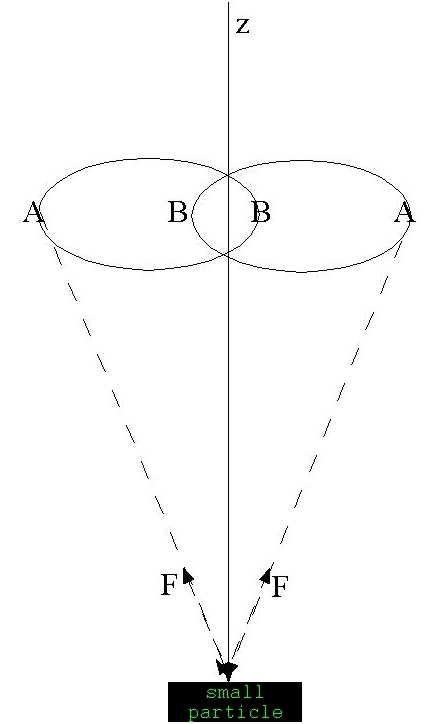}
\end{center}
\caption{Sitnikov problem, two large masses following keplerian orbits and a small particle which can attain large velocity away from the system.}
\label{Rys.1}
\end{figure}

\section{Conclusion}

It is almost 350 years since Newton formulated his famous laws of motion and gravity. He quickly showed how to solve they for two-body problem but his attempts to solve three-body problem failed. Since this time the brightest men tried to solve it but they were unsuccesful too. Apart from some special solution (like Lagrange one, 1772) no general exist \cite{wie,roy,str}. In this paper we proposed a special 3-body case known as a Sitnikov problem as an explanation of jets in astrophysical systems. This effect can also be used in future interstellar travels, Sitnikov problem playing a role of a gravitational assist.

\section{Appendix}
The program was written using Quincy2005 (MinGW gcc 4.2.1) and executed under Windows XP on Pentium M processor (IBM ThinkPad X40 machine). To save the results in a file run: program.exe$>$sitnikov.txt

\lstset{language=C++,basicstyle=\footnotesize}
\begin{lstlisting}
	#include<iostream>
	#include<cmath>
	using namespace std;
	const double pi=3.1415926;
	double x,r,z,v,a,b,GM,okres,e,t,dt,
	k1z,k2z,k3z,k4z,k1v,k2v,k3v,k4v;

	double keplereq (double t);

	int main()
	{a=6e+07; b=2e+07;
	 GM=1e+22;//each mass
	 e=sqrt(1-b*b/(a*a));//eccentricity
	 okres=2*pi*pow(a,1.5)/sqrt(2*GM);
	 dt=0.01; t=0; z=-4.931*a; v=0; r=a*(1-e)/2;
	
	 cout <<"a= "<<a<<"  GM= "<<GM<<"  e= "<<e<<"  z0= "<<z<<"\n\n";
	 
	  for (t=0; t<=100; t=t+dt)
	    {k1z=v*dt;
	     k1v=-2*GM*z/pow((z*z+r*r),1.5)*dt;
 	     k2z=(v+0.5*k1v)*dt;
	     k2v=-2*GM*(z+0.5*k1z)/pow(((z+0.5*k1z)*(z+0.5*k1z)+r*r),1.5)*dt;
	     k3z=(v+0.5*k2v)*dt;
	     k3v=-2*GM*(z+0.5*k2z)/pow(((z+0.5*k2z)*(z+0.5*k2z)+r*r),1.5)*dt;
	     k4z=(v+0.5*k1v)*dt;
	     k4v=-2*GM*(z+k3z)/pow(((z+k3z)*(z+k3z)+r*r),1.5)*dt;

 	     z=z+(k1z+2*k2z+2*k3z+k4z)/6;
 	     v=v+(k1v+2*k2v+2*k3v+k4v)/6;
 	     r=a*(1-e*cos(keplereq(t)))/2;
	     cout << "t= "<<t<<"   z= "<<z<<"   v= "<<v<<"   r= "<<r<<"\n";
	    }
	}

	double keplereq (double t)
	{double f,fmin,u,umin;
	 fmin=100; f=0; u=0; umin=0;
	 if (t>=okres)
	   {while (t>=okres) {t=t-okres;}}
	 for (u=0;u<2*pi;u=u+0.01)
	   {f=u-e*sin(u)-2*pi*t/okres;
	    if (fabs(f) < fabs(fmin))
	      {fmin=f;umin=u;}
	   } 
	 return umin;
	}
\end{lstlisting}

\end{document}